\documentclass[twocolumn,showpacs,amsfonts,amsmath,amssymb,aps,pra,floatfix]{revtex4}
\usepackage{graphicx}
\usepackage{bm}
\begin{document}
\title{Formulation of the uncertainty relations in terms of the R\'enyi entropies}
\author{Iwo Bialynicki-Birula}
\email{birula@cft.edu.pl} \affiliation{Center for Theoretical Physics, Polish Academy of Sciences\\Al. Lotnik\'ow 32/46, 02-668 Warsaw, Poland}

\begin{abstract}
Quantum-mechanical uncertainty relations for position and momentum are expressed in the form of inequalities involving the R\'enyi entropies. The proof of these inequalities requires the use of the exact expression for the $(p,q)$-norm of the Fourier transformation derived by Babenko and Beckner. Analogous uncertainty relations are derived for angle and angular momentum and also for a pair of complementary observables in $N$-level systems. All these uncertainty relations become more attractive when expressed in terms of the symmetrized R\'enyi entropies.
\end{abstract}
 \maketitle

\section{Introduction}

The R\'enyi entropy is a one-parameter extension of the Shannon entropy. There is extensive literature on the applications of the R\'enyi entropy in many fields from biology, medicine, genetics, linguistics, and economics to electrical engineering, computer science, geophysics, chemistry, and physics. My aim is to describe the limitations on the information characterizing quantum systems, in terms of the R\'enyi entropies. These limitations have the from of inequalities that have the physical interpretation of the uncertainty relations.

The R\'enyi entropy has been widely used in the study of quantum systems. In particular, it was used in the analysis of quantum entanglement \cite{gl,asi,bcehas,ter,bez}, quantum communication protocols \cite{rgk,gl1}, quantum correlations \cite{lnp}, quantum measurement \cite{bg}, and decoherence \cite{kh}, multiparticle production in high-energy collisions \cite{bc,bcz,bcz1}, quantum statistical mechanics \cite{mw}, localization properties of Rydberg states \cite{arb} and spin systems \cite{gz,vc}, in the study of the quantum-classical correspondence \cite{dmfs}, and the localization in phase space \cite{vp}. In view of these numerous and successful applications, it seems worthwhile to formulate the quantum-mechanical uncertainty relations for canonically conjugate variables in terms of the R\'enyi entropies. I do not want to enter here into the discussion (cf. \cite{bz,cgt}) of a fundamental problem: which (if any) entropic measure of uncertainty is adequate in quantum mechanical measurements. The uncertainty relations derived in this paper are valid as mathematical inequalities, regardless of their physical interpretation.

The R\'enyi entropy is defined \cite{renyi} as
\begin{eqnarray}\label{renyi}
H_\alpha=\frac{1}{1-\alpha}\ln\left(\sum p_k^\alpha\right).
\end{eqnarray}
R\'enyi called this quantity ``the measure of information of order $\alpha$ associated with the probability distribution ${\cal P}=(p_1,\dots p_n)$''. The R\'enyi measure of information $H_\alpha$ may also be viewed as a measure of uncertainty since, after all, the uncertainty is the missing information. In the formulation of the uncertainty relations given below, the R\'enyi entropies will be used as the measures of uncertainties.

In order to simplify the derivations, I use the natural logarithm in the definition (\ref{renyi}) of the R\'enyi entropy. However, all uncertainty relations derived in this paper [Eqs.~(\ref{reur}), (\ref{reurn}), (\ref{xpsym}), (\ref{angle}), (\ref{eurn}), (\ref{bbmr}), and (\ref{bbmr1})] have the same form for all choices of the base of the logarithm because they are homogeneous in $\ln(\dots)$. Note that the definition of the R\'enyi entropy is also applicable when the sum has infinitely many terms, provided this infinite sum converges. The R\'enyi entropy (\ref{renyi}) is a nonincreasing function of $\alpha$ \cite{renyi}. For $\alpha >\beta$ we have $H_\alpha\leq H_\beta$. In the limit, when $\alpha\to 1$ the R\'enyi entropy is equal (apart from a different base of the logarithm) to the Shannon entropy
\begin{eqnarray}\label{shan}
\lim_{\alpha\to 1} H_\alpha=-\sum p_k\ln p_k.
\end{eqnarray}

According to the probabilistic interpretation of quantum theory, the probability distribution associated with the measurement of a physical variable represented by the operator $A$ is defined as
\begin{eqnarray}\label{prob}
p_k={\rm Tr}\{\rho P_k\},
\end{eqnarray}
where $\rho$ is the density operator describing the state of the quantum system, and $P_k$ is the projection operator corresponding to the $k$th segment of the spectrum of $A$ (the $k$th bin). The uncertainty is the lowest when only one $p_k$ is different from zero --- the R\'enyi entropy reaches then its lowest value: zero.

The probability distributions $p_k^A$ and $p_k^B$  that correspond to different physical variables but to the same state of the system are, in general, correlated. These correlations lead to restrictions on the values of the R\'enyi entropies $H_\alpha^A$ and $H_\beta^B$. When these restrictions have the form of an inequality $H_\alpha^A+H_\beta^B\geq C>0$, they deserve the name of the uncertainty relations because not only do they prohibit the vanishing of {\it both} uncertainties for the same state but they also require that one uncertainty must increase when the other decreases.

In the present paper, I derive the inequalities for three pairs of observables: position and momentum (or time and frequency), angle and angular momentum, and the complementary observables --- the analogs of $x$ and $p$ --- in finite dimensional spaces. These inequalities are generalizations of the entropic uncertainty relations established before for the Shannon entropies \cite{d,p,ibb}. There is some overlap in mathematical derivations [especially in the extensive use of $(p,q)$-norms] between the results presented in this paper and the earlier works of Maassens and Uffink \cite{mu,u} and Rajagopal \cite{raja}. However, these authors did not express the uncertainty relations in terms of the R\'enyi entropies and they did not introduce the finite resolutions that characterize all physical measurements.

\section{Uncertainty relations for $x$ and $p$}

The probability distributions associated with the measurements of momentum and position of a quantum particle in a pure state (generalization to mixed states will be given in Sec.~\ref{mixed}) are
\begin{eqnarray}\label{pos_mom}
p_k=\int_{k\delta p}^{(k+1)\delta p}\!\!\!dp\,\vert{\tilde\psi}(p)\vert^2,\;\;q_l=\int_{l\delta x}^{(l+1)\delta x}\!\!\!dx\vert\psi(x)\,\vert^2,
\end{eqnarray}
where I have assumed that the sizes of all bins are the same. The indices $k$ and $l$ run from $-\infty$ to $\infty$ and the Fourier transform is defined with the physical normalization, i.e.
\begin{eqnarray}\label{ft}
{\tilde{\psi}}(p)=\frac{1}{\sqrt{2\pi\hbar}}\int\!dx\,e^{- ipx/\hbar}\psi(x).
\end{eqnarray}
From the two probability distributions (\ref{pos_mom}) we may construct the R\'enyi entropies $H^{(p)}_{\alpha}$ and $H^{(x)}_{\beta}$ that measure the uncertainty in momentum and position
\begin{eqnarray}\label{renyixp}
H^{(p)}_\alpha=\frac{1}{1-\alpha}\ln\!\left(\sum p_k^\alpha\right),H^{(x)}_{\beta}=\frac{1}{1-\beta}\ln\!\left(\sum q_l^\beta\right).
\end{eqnarray}

I shall prove in the next section that the uncertainty relation restricting the values of $H^{(p)}_{\alpha}$ and $H^{(x)}_{\beta}$ has the following form:
\begin{eqnarray}\label{reur}
H^{(p)}_\alpha+H^{(x)}_\beta\geq-\frac{1}{2}\left(\frac{\ln \alpha}{1-{\alpha}}+\frac{\ln \beta}{1-\beta}\right)-\ln\left(\frac{\delta x\delta p}{\pi\hbar}\right),
\end{eqnarray}
where the parameters $\alpha$ and $\beta$ are assumed to be positive and they are constrained by the relation
\begin{eqnarray}\label{relation}
\frac{1}{\alpha}+\frac{1}{\beta}=2.
\end{eqnarray}
In the limit, when $\alpha\to 1$ and ${\beta}\to 1$, this uncertainty relation reduces to the uncertainty relation for the Shannon entropies
\begin{eqnarray}\label{seur}
H^{(p)} + H^{(x)}\geq -\ln\left(\frac{\delta x\delta p}{e\pi\hbar}\right)
\end{eqnarray}
that had already been derived some time ago \cite{ibb}.

Note that the relations (\ref{reur}) and (\ref{seur}) are quite different from the standard uncertainty relations. As has been aptly stressed by Peres \cite{ap}, ``The uncertainty relation such as $\Delta x\Delta p\geq\hbar/2$ is not a statement about the accuracy of our measuring instruments.'' In contrast, both entropic uncertainty relations (\ref{reur}) and (\ref{seur}) {\it do depend} on the accuracy of the measurement --- they explicitly contain the area of the phase-space $\delta x\delta p$ determined by the resolution of the measuring instruments. This aspect of the uncertainty relations (\ref{reur}) and (\ref{seur}) can be summarized as follows: the more precisely one wants to localize the particle in the phase space, the larger the sum of the uncertainties in $x$ and $p$.

The uncertainty relation (\ref{reur}) is not sharp --- its improvement is a challenging open problem. However, it becomes sharper and sharper when the relative size of the phase space area ${\delta x\delta p}/(\pi\hbar)$ defined by the experimental resolutions decreases, as it is when we enter deeper and deeper into the quantum regime.

\section{Proof}

The proof of the inequality (\ref{reur}) employs the known value of the $(p,q)$-norm of the Fourier transformation. The $(p,q)$-norm of an operator $T$ is defined as the smallest number $k(p,q)$ such that for all $\psi$
\begin{eqnarray}\label{pq}
\|{T{\psi}}\|_p\leq k(p,q)\,\|{\psi}\|_{q},
\end{eqnarray}
where the $p$-norm (or the $q$-norm) of a function is defined in the standard way
\begin{eqnarray}\label{norm}
\|{\psi}\|_p=\left(\int_{-\infty}^\infty\!dx|{\psi(x)}|^p\right)^{1/p},
\end{eqnarray}
and the values of the parameters $p$ and $q$ satisfy the conditions
\begin{eqnarray}\label{cond}
\frac{1}{p}+\frac{1}{q}=1,\;\;p\geq q.
\end{eqnarray}
The parameters $p$ and $q$ should no be confused with momentum and position.

The $(p,q)$-norm of the Fourier transformation has been found for even values of $p$ by Babenko \cite{bab} and for all values of $p$ by Beckner \cite{beck}. For the physical normalization (\ref{ft}) of the Fourier transform, the Babenko-Beckner inequality reads
\begin{eqnarray}\label{bb}
\|{\tilde{\psi}}\|_p\leq k(p,q)\|{\psi}\|_{q},
\end{eqnarray}
where
\begin{eqnarray}\label{k}
k(p,q)=\left(\frac{p}{2\pi\hbar}\right)^{-\frac{1}{2p}} \left(\frac{q}{2\pi\hbar}\right)^{\frac{1}{2q}}.
\end{eqnarray}
Since the function $\psi$ can be treated as the Fourier transform of $\tilde{\psi}$, the following inequality also holds:
\begin{eqnarray}\label{bb1}
\|\psi\|_p\leq k(p,q)\|{\tilde{\psi}}\|_{q}.
\end{eqnarray}
The inequalities (\ref{bb}) and (\ref{bb1}) are saturated by all Gaussian functions.

In terms of the probability densities $\tilde{\rho}(p)=|\tilde{\psi}(p)|^2$ and $\rho(x)=|{\psi(x)}|^2$, the inequalities (\ref{bb}) and (\ref{bb1}) read
\begin{subequations}\label{bbph}
\begin{eqnarray}
\left(\int_{-\infty}^\infty\!\!\!\!dp\, [{\tilde{\rho}(p)}]^{\alpha}\right)^{\frac{1}{\alpha}}\!\leq n(\alpha,\beta)\left(\int_{-\infty}^\infty\!\!\!\!dx\, [{\rho}(x)]^{\beta}\right)^{\frac{1}{\beta}}\!\!,\label{bbpha}\\
\left(\int_{-\infty}^\infty\!\!\!\!dx\, [{\rho}(x)]^{\alpha}\right)^{\frac{1}{\alpha}}\!\leq n(\alpha,\beta)\left(\int_{-\infty}^\infty\!\!\!\!dp\, [{\tilde{\rho}(p)}]^{\beta}\right)^{\frac{1}{\beta}}\!\!,\label{bbphb}
\end{eqnarray}
\end{subequations}
where $\alpha=p/2$, $\beta=q/2$, $\alpha\geq\beta$, and
\begin{eqnarray}\label{n}
n(\alpha,\beta)= \left(\frac{\alpha}{\pi\hbar}\right)^{-\frac{1}{2{\alpha}}} \left(\frac{\beta}{\pi\hbar}\right)^{\frac{1}{2\beta}}.
\end{eqnarray}

In the first part of the proof I shall use the inequality (\ref{bbpha}). In order to relate this inequality to the R\'enyi entropies (\ref{renyixp}), I shall first split the full integration ranges into the $\delta p$ and $\delta x$ bins
\begin{subequations}\label{split}
\begin{eqnarray}
\int_{-\infty}^\infty dp[\tilde{\rho}(p)]^{\alpha} = \sum_{k=-\infty}^\infty\int_{k\delta
p}^{(k+1)\delta p}\!\!\!dp\,[\tilde{\rho}(p)]^{\alpha},\\
\int_{-\infty}^\infty dx[{\rho}(x)]^{\beta} = \sum_{l=-\infty}^\infty\int_{l\delta x}^{(l+1)\delta x}\!\!\!dx\,[{\rho}(x)]^{\beta}.
\end{eqnarray}
\end{subequations}
Next, for each term in these sums I shall use the integral form of the Jensen inequality \cite{hlp,ms}. For convex functions this inequality can be stated as follows: the value of the function at the average point does not exceed the average value of the function. For concave functions it is just the opposite: the average value of the function does not exceed the value of the function at the average point. Since for $\alpha>1$ the function $f(z)= z^\alpha$ is convex and for $\beta<1$ the function $g(z)= z^\beta$ is concave, we obtain the following two inequalities:
\begin{subequations}\label{jensen}
\begin{eqnarray}
\left(\frac{1}{\delta p}\int_{k\delta p}^{(k+1)\delta p}\!\!\!dp\,\tilde{\rho}(p)\right)^{\!{\alpha}}\leq \frac{1}{\delta p}\int_{k\delta
p}^{(k+1)\delta p}\!\!\!dp\,[\tilde{\rho}(p)]^{{\alpha}},\;\;\\
\frac{1}{\delta x}\int_{l\delta x}^{(l+1)\delta x}\!\!\! dx[{\rho}(x)]^{\beta}\leq \left(\frac{1}{\delta x}\int_{l\delta x}^{(l+1)\delta x}\!\!\!dx\,{\rho}(x)\right)^{\!\beta}.\;\;\;
\end{eqnarray}
\end{subequations}
Therefore, with the use of the definitions of the probabilities (\ref{pos_mom}), we may convert Eqs.~(\ref{split}) into the following inequalities:
\begin{subequations}\label{jensen1}
\begin{eqnarray}
(\delta p)^{1-\alpha}\sum_{l=-\infty}^\infty p_k^\alpha \leq
\int_{-\infty}^\infty\!\!\!dp\,[\tilde{\rho}(p)]^{\alpha},\\
\frac{1}{\delta x}\int_{-\infty}^\infty\!\!\! dx[{\rho}(x)]^{\beta}\leq (\delta x)^{1-\beta}\sum_{l=-\infty}^\infty q_l^{\beta}.
\end{eqnarray}
\end{subequations}
These inequalities combined with the Babenko-Beckner result (\ref{bbpha}) give
\begin{eqnarray}\label{bbph1}
\left(\!(\delta p)^{1-\alpha}\!\!\!\sum_{k=-\infty}^\infty\! p_k^\alpha\!\right)^{\!\frac{1}{\alpha}}\!\!\!\leq n(\alpha,\beta)\left(\!(\delta x)^{1-\beta}\!\!\!\sum_{l=-\infty}^\infty\! q_l^\beta\!\right)^{\!\frac{1}{\beta}}\!\!.
\end{eqnarray}
This inequality does not depend on the choice of units used to measure $\delta x$, $\delta p$, and $\hbar$ since it can be transformed to the following dimensionless form
\begin{eqnarray}\label{bbph2}
\left(\sum_{k=-\infty}^\infty\! p_k^\alpha\!\right)^{\!\frac{1}{\alpha}}\!\!\!\leq \gamma^{\frac{1}{2\beta}-\frac{1}{2\alpha}}\left(\frac{\alpha}{\pi}\right)^{-\frac{1}{2{\alpha}}} \left(\frac{\beta}{\pi}\right)^{\frac{1}{2\beta}}\!\!\left(\sum_{l=-\infty}^\infty\! q_l^\beta\!\right)^{\!\frac{1}{\beta}}\!\!,
\end{eqnarray}
where $\gamma=\delta x\delta p/\hbar$. After raising both sides of this inequality to the (positive) power $\alpha/(\alpha-1)=\beta/(1-\beta)$ and with the use of the relation $1/(1-\alpha)+1/(1-\beta)=2$, we obtain
\begin{eqnarray}\label{bbph3}
\left(\,\sum_{k=-\infty}^\infty\! p_k^{\alpha}\!\right)^{\!\frac{1}{\alpha-1}}\!\!\!\leq \frac{\delta x\delta p}{\pi\hbar}\,\frac{\beta^{\frac{1}{2(1-\beta)}}} {{\alpha}^{\frac{1}{2(\alpha-1)}}} \left(\,\sum_{l=-\infty}^\infty\! q_l^{\beta}\!\right)^{\!\frac{1}{1-\beta}}\!\!.
\end{eqnarray}
Finally, by taking the logarithm of both sides we obtain the uncertainty relation (\ref{reur}) but only for $\alpha>\beta$. To extend this result to the values $\alpha<\beta$, we have to start from the inequality (\ref{bbphb}) instead of (\ref{bbpha}).

In order to generalize these results to $n$ dimensions we need the following value of the $(p,q)$-norm for the $n$-dimensional Fourier transform \cite{beck}
\begin{eqnarray}\label{kn}
k_n(p,q)=\left(\frac{p}{2\pi\hbar}\right)^{-\frac{n}{2p}} \left(\frac{q}{2\pi\hbar}\right)^{\frac{n}{2q}}.
\end{eqnarray}
The uncertainty relations are then obtained in the same way as in the one-dimensional case and they have the form:
\begin{eqnarray}\label{reurn}
H^{(p)}_\alpha+H^{(x)}_{\beta}\geq-\frac{n}{2}\!\left(\!\frac{\ln \alpha}{1-{\alpha}}+\frac{\ln \beta}{1-\beta}\!\right)-n \ln\!\left(\!\frac{\delta x\delta p}{\pi\hbar}\!\right)\!.
\end{eqnarray}

\section{Uncertainty relations for $\varphi$ and $M_z$}

The uncertainty relations in terms of the R\'enyi entropies can also be formulated for the angle $\varphi$ and the angular momentum $M_z$ and they have the form
\begin{align}\label{angle}
H_\alpha^{(M_z)} + H_\beta^{(\varphi)}\geq -\ln\frac{\delta\varphi}{2\pi}.
\end{align}
The probability distributions $p_m^{(M_z)}$ and $p_l^{(\varphi)}$ that are used to calculate these R\'enyi entropies are defined as follows:
\begin{align}\label{angle1}
 p_m^{(M_z)}=|c_m|^2,\;\;\;p_l^{(\varphi)}= \int_{l \delta\varphi}^{(l+1)\delta\varphi}\!\!\!d\varphi
|\psi(\varphi)|^2,
\end{align}
where the amplitudes $c_m$ are the coefficients in the expansion of $\psi(\varphi)$ into the eigenstates of $M_z$,
\begin{eqnarray}\label{standard}
\psi(\varphi)=\frac{1}{\sqrt{2\pi}}\sum_{m=-\infty}^\infty c_m e^{im\varphi},
\end{eqnarray}
and $\delta\varphi$ is the experimental resolution in the measurement of the angular distribution. In contrast to the uncertainty relation for position and momentum, the inequality (\ref{angle}) is sharp (it is saturated by any eigenstate of $M_z$) and the bound does not depend on $\alpha$ and $\beta$. The absence of the Planck in this uncertainty relation is due to a cancellation --- the volume of the phase space defined by the experimental resolution is $\delta\varphi\,\delta M_z=\delta\varphi\hbar$ and the standard reference volume in quantum theory is $2\pi\hbar$.

The proof of this inequality can be obtained along similar lines as the proof of its counterpart for $x$ and $p$ but the starting point is now two Young-Hausdorff inequalities for the Fourier series \cite{hlp,zyg}
\begin{subequations}
\begin{eqnarray}\label{hy}
\left(\sum_{m=-\infty}^\infty |c_m|^p\right)^\frac{1}{p}\leq l(p,q)\left(\int_0^{2\pi}\!d\varphi|\psi(\varphi)|^q\right)^\frac{1}{q}\!,\\
\left(\int_0^{2\pi}\!d\varphi|\psi(\varphi)|^p\right)^\frac{1}{p}\leq l(p,q)\left(\sum_{m=-\infty}^\infty |c_m|^q\right)^\frac{1}{q}\!,
\end{eqnarray}
\end{subequations}
where
\begin{align}\label{anorm}
l(p,q)=(2\pi)^{\frac{1}{2p}-\frac{1}{2q}}.
\end{align}
Choosing either the first or the second inequality, we obtain the inequality (\ref{angle}) either for $\alpha>\beta$ or for $\alpha<\beta$.

The uncertainty relations (\ref{angle}) and (\ref{sangle}) also hold for the phase and the occupation number of a mode of radiation. In this case, the Fourier expansion (\ref{standard}) contains only the terms with $m\geq 0$.

\section{Uncertainty relations for $N$-level systems}

For quantum systems described by vectors in the $N$-dimensional Hilbert space the analog of the uncertainty relation for the R\'enyi entropies is
\begin{eqnarray}\label{eurn}
\frac{1}{1-\alpha}\ln\left(\sum_{k=1}^N \tilde{\rho}_k^\alpha\right) +\frac{1}{1-\beta}\ln\left(\sum_{l=1}^N \rho_l^\beta\right)\geq \ln N,
\end{eqnarray}
where $\tilde{\rho}_k=|\tilde{a}_k|^2,\;\rho_l=|{a}_l|^2$ and the amplitudes $\tilde{a}_k$ and ${a}_l$ are connected by the discrete Fourier transformation
\begin{eqnarray}\label{discf}
\tilde{a}_k=\frac{1}{\sqrt{N}}\sum_{l=1}^N \exp(2\pi i k\,l/N)\,a_l.
\end{eqnarray}
The complex numbers $\tilde{a}_k$ and ${a}_l$ can be interpreted as the probability amplitudes to find a particle in the discretized momentum space and position space \cite{kraus}, but they can also be viewed as amplitudes in a general abstract $N$-dimensional Hilbert space. The uncertainty relation (\ref{eurn}) is saturated for the states that are localized either in ``position space'' (only one of the amplitudes $a_l$ is different from zero) or in ``momentum space'' (only one of the amplitudes $\tilde{a}_k$ is different from zero). Like in the case of the uncertainty relations for the angle and the angular momentum, the bound does not depend on $\alpha$ and $\beta$. The absence of the Planck constant is again due to a cancellation --- it would reappear if $l$ and $k$ in (\ref{discf}) are given the physical dimension of length and momentum.

The proof of the uncertainty relation (\ref{eurn}) proceeds along similar lines as the proof of (\ref{reur}) but now we invoke a different known inequality --- the $(p,q)$-norm of the discrete Fourier transform (cf., for example, Ref.~\cite{rs})
\begin{eqnarray}\label{discf1}
\|\tilde{a}\|_p\leq N^{\frac{1}{2p}-\frac{1}{2q}}\|a\|_q,\;\;\|a\|_p\leq N^{\frac{1}{2p}-\frac{1}{2q}}\|\tilde{a}\|_q.
\end{eqnarray}
Uncertainty relations for $N$-level systems involving the $(p,q)$-norms of the discrete Fourier transform were established in Ref.~\cite{kp} but they have not been used to derive the uncertainty relations for the R\'enyi entropies.

For a system composed of two subsystems described by state vectors in the $N$ and $M$ dimensional spaces the bound on the right hand side of the inequality is equal to $\ln(N M)=\ln N+\ln M$ because the dimensionality of the Hilbert space of the composed system is $N M$. The same result is obtained for two totally independent systems of dimensionality $N$ and $M$ because the R\'enyi entropy is additive for independent probability distributions. This means that the uncertainty relation is already saturated by separable states and allowing for entanglement does not make any difference.

\section{Uncertainty relations for mixed states}\label{mixed}

The uncertainty relations for the R\'enyi entropies hold also for all mixed states. This result is not obvious because the R\'enyi entropy is not a convex function \cite{bez} of the probability distributions for all values of $\alpha$. Hence, the terms on the left-hand side of the uncertainty relation (\ref{reur}) may decrease as a result of mixing. However, I shall prove now that the inequalities (\ref{bbph}) that were the starting point in the derivations hold also for mixed states. This follows from the integral form of the Minkowski inequalities \cite{bb}, namely
\begin{subequations}\label{mink}
\begin{eqnarray}
\left(\int\!dV\, |f+g|^{\alpha}\right)^{\!\frac{1}{\alpha}}\!\leq \left(\int\!dV\, |f|^{\alpha}\right)^{\!\frac{1}{\alpha}}\!+\left(\int\!dV\, |g|^{\alpha}\right)^{\!\frac{1}{\alpha}}\!\!,\nonumber\\
\\
\left(\int\!dV\, |f|^{\beta}\right)^{\!\frac{1}{\beta}}\!+\left(\int\!dV\, |g|^{\beta}\right)^{\!\frac{1}{\beta}}\leq \left(\int\!dV\, |f+g|^{\beta}\right)^{\!\frac{1}{\beta}}\!\!,\nonumber\\
\end{eqnarray}
\end{subequations}
where $f$ and $g$ are nonnegative functions and the parameters $\alpha$ and $\beta$ satisfy the condition $\alpha>\beta$. Substituting in the first inequality for the functions $f$ and $g$ the weighted densities in momentum space $f=\lambda{\tilde{\rho}_1(p)}$ and $g=(1-\lambda){\tilde{\rho}_2(p)}$ and in the second inequality the weighted densities in the coordinate space $f=\lambda\rho_1(x)$ and $g=(1-\lambda)\rho_2(x)$, we obtain
\begin{subequations}\label{mink1}
\begin{align}\label{mink1a}
&\left(\int_{-\infty}^\infty\!\!\!dp\, (\lambda\tilde{\rho}_1(p)+(1-\lambda)\tilde{\rho}_2(p))^{\alpha}\right)^{\frac{1}{\alpha}}\nonumber\\
&\leq \lambda\left(\int_{-\infty}^\infty\!\!\!dp\,\tilde{\rho}_1(p))^{\alpha}\right)^{\frac{1}{\alpha}}\!\!
+(1-\lambda)\left(\int_{-\infty}^\infty\!\!\!dp\,\tilde{\rho}_2(p))^{\alpha}\right)^{\frac{1}{\alpha}},\\
&\lambda\left(\int_{-\infty}^\infty\!\!\!dx\,\rho_1(x))^{\beta}\right)^{\frac{1}{\beta}}\!\!
+(1-\lambda)\left(\int_{-\infty}^\infty\!\!\!dx\,\rho_2(x))^{\beta}\right)^{\frac{1}{\beta}}\nonumber\\
&\leq\left(\int_{-\infty}^\infty\!\!\!dx\, (\lambda\rho_1(x)+(1-\lambda)\rho_2(x))^{\beta}\right)^{\frac{1}{\beta}}.
\end{align}
\end{subequations}
Comparing these results with the weighted sum of inequalities (\ref{bbpha}) for pure states
\begin{align}\label{bbphm}
&\lambda\left(\int_{-\infty}^\infty\!\!\!dp\,({\tilde{\rho}_1(p)})^{\alpha}\right)^{\frac{1}{\alpha}} +(1-\lambda)\left(\int_{-\infty}^\infty\!\!\!dp\,
({\tilde{\rho}_2(p)})^{\alpha}\right)^{\frac{1}{\alpha}}\nonumber\\
&\leq n(\alpha,\beta)\lambda\left(\int_{-\infty}^\infty\!\!\!dx\, (\lambda{\rho}_1(x))^{\beta}\right)^{\frac{1}{\beta}}\nonumber\\
&+n(\alpha,\beta)(1-\lambda)\left(\int_{-\infty}^\infty\!\!\!dx\, (\lambda{\rho}_2(x))^{\beta}\right)^{\frac{1}{\beta}},
\end{align}
we extend the inequality (\ref{bbpha}) to mixed states
\begin{align}\label{bbphm1}
&\left(\int_{-\infty}^\infty\!\!\!dp\, (\lambda\tilde{\rho}_1(p)+(1-\lambda)\tilde{\rho}_2(p))^{\alpha}\right)^{\frac{1}{\alpha}}\nonumber\\
&\leq n(\alpha,\beta)\left(\int_{-\infty}^\infty\!\!\!dx\, (\lambda\rho_1(x)+(1-\lambda)\rho_2(x))^{\beta}\right)^{\frac{1}{\beta}}\!\!.
\end{align}
In the same manner we can extend the inequality (\ref{bbphb}) to mixed states.

Once we have proven the validity of the inequalities (\ref{bbph}) for mixed states, we may proceed as before to prove the validity of the the R\'enyi uncertainty relations (\ref{reur}) for mixed states. Similar arguments can be invoked to prove also the uncertainty relations (\ref{angle}) and (\ref{eurn}) for mixed states.

\section{Uncertainty relations for continuous distributions}

There exist also purely mathematical versions of the uncertainty relations that {\it do not involve} the experimental resolutions $\delta x$ and $\delta p$ of the measuring devices. By taking directly the logarithm of the inequality (\ref{bbph}), and using the relations between $\alpha$ and $\beta$, we arrive at
\begin{align}\label{bbphc}
\frac{1}{1-\alpha}\ln &\left(\int_{-\infty}^\infty\!\!\!dp\,[{\tilde{\rho}(p)}]^{\alpha}\right) +\frac{1}{1-\beta}\ln\left(\int_{-\infty}^\infty\!\!\!dx\,[{\rho}(x)]^{\beta}\right)\nonumber\\
&\geq -\frac{1}{2(1-\alpha)}\ln\frac{\alpha}{\pi} -\frac{1}{2(1-\beta)}\ln\frac{\beta}{\pi}.
\end{align}
On the left-hand side of this inequality we have what might be called the continuous or integral versions of the R\'enyi entropies. To derive this mathematical inequality, I have dropped $\hbar$ in the definition (\ref{ft}) of the Fourier transform. This inequality has been also recently independently proven by Zozor and Vignat \cite{zv}. Analogous relations for the continuous Tsallis entropies for $x$ and $p$ were obtained by Rajagopal \cite{raja}.

In the limit, when $\alpha\to 1,\;\beta\to 1$, we obtain from the inequality (\ref{bbphc}) the entropic uncertainty relation in the form
\begin{eqnarray}\label{bbmb}
-\int_{-\infty}^\infty\!\!\!\!dp\,{\tilde{\rho}(p)}\ln{\tilde{\rho}(p)} -\int_{-\infty}^\infty\!\!\!\!dx\,{\rho}(x)\ln{\rho}(x)\geq \ln(e\pi)
\end{eqnarray}
that had been conjectured by Hirschman \cite{hirsch} and later proved by Bialynicki-Birula and Mycielski \cite{bbm} and by Beckner \cite{beck}. The inequalities (\ref{bbphc}) and (\ref{bbmb}) are saturated by the Gaussian probability distributions.

For wave functions defined over an $n$-dimensional space, the bound on the right-hand side in (\ref{bbphc}) and (\ref{bbmb}) is just multiplied by $n$, as in the previous formula (\ref{reurn}). Therefore also in this case, like in the finite-dimensional case, the uncertainty relations are already saturated by separable states.

In a similar fashion we can derive the uncertainty relation for $\varphi$ and $M_z$ that does not involve the resolution $\delta\varphi$.
\begin{align}\label{bbmr}
\frac{1}{1-\alpha}\ln \left(\sum_{-\infty}^\infty \rho_m^{\alpha}\right) &+\frac{1}{1-\beta}\ln\left(\int_{0}^{2\pi}\!\!\!d\varphi\,[{\rho}(\varphi)]^{\beta}\right)\nonumber\\
&\geq \ln(2\pi),
\end{align}
where $\rho_m=|c_m|^2$ and ${\rho}(\varphi)=|\psi(\varphi)|^2$. In the limit, when $\alpha\to 1$ and $\beta\to 1$, we obtain the mathematical entropic uncertainty relation for the angle and the angular momentum derived before \cite{bbm}
\begin{eqnarray}\label{bbmr1}
-\sum_{-\infty}^\infty \rho_m \ln \rho_m-\int_{0}^{2\pi}\!\!\!\!d\varphi\,{\rho}(\varphi) \ln {\rho}(\varphi)\geq \ln(2\pi).
\end{eqnarray}
The inequalities (\ref{bbmr}) and (\ref{bbmr1}), like their discrete counterpart (\ref{angle}), are saturated when the Fourier series (\ref{standard}) has only one term.

\section{Symmetrized R\'enyi entropies}

In the uncertainty relations for the R\'enyi entropies the parameters $\alpha$ and $\beta$ appear always in conjugate pairs. This observation suggests the introduction of the {\em symmetrized R\'enyi entropy} ${\cal H}_s$ defined as follows
\begin{eqnarray}\label{sym}
{\cal H}_s=\frac{1}{2}\left(H_\alpha+H_\beta\right),
\end{eqnarray}
where $\alpha$ and $\beta$ satisfy the conditions (\ref{cond}) and they are related to the parameter $s$ through the formulas
\begin{eqnarray}\label{form}
\alpha=\frac{1}{1-s},\;\;\;\beta=\frac{1}{1+s},\;\;-1\leq s\leq 1.
\end{eqnarray}
The symmetrized R\'enyi entropy ${\cal H}_s$ is a symmetric function of $s$ and for $s=0$ it becomes the Shannon entropy. The uncertainty relations expressed in terms of the symmetrized R\'enyi entropies have the form
\begin{align}\label{xpsym}
&{\cal H}^{(p)}_s+{\cal H}^{(x)}_s\nonumber\\
&\geq\frac{1}{2}\left(\ln(1-s^2)+\frac{1}{s}\ln\frac{1+s}{1-s}\right)-\ln\left(\frac{\delta x\delta p}{\pi\hbar}\right).
\end{align}
They are obtained by taking half of the sum of the inequality (\ref{reur}) and the inequality obtained from (\ref{reur}) by interchanging $\alpha$ and $\beta$. The same symmetrization procedure can be applied to all other uncertainty relations derived in this paper. In particular, we obtain
\begin{align}\label{sangle}
{\cal H}_s^{(M_z)} + {\cal H}_s^{(\varphi)}\geq -\ln\frac{\delta\varphi}{2\pi}.
\end{align}
Analogous symmetrized versions of the uncertainty relations for the Tsallis entropies were introduced also by Rajagopal \cite{raja}.

In contrast to the inequalities that contain the R\'enyi entropies $H_\alpha$ and $H_\beta$, in the uncertainty relations that contain the symmetrized entropy the same measure of uncertainty is used for both physical variables. This is clearly a desirable feature but it remains to be seen whether the symmetrized R\'enyi entropy (\ref{sym}) is a useful concept outside the realm of the uncertainty relations.

Different uncertainty relations in which the same measure of uncertainty is used for both variables follow from the fact that the R\'enyi entropy is a nonincreasing function of $\alpha$. For example, for the position and momentum we obtain
\begin{align}\label{xpsym1}
&H^{(p)}_\beta+H^{(x)}_\beta\nonumber\\
&\geq-\ln \beta-\frac{\beta-1/2}{1-\beta}\ln(2\beta-1)-\ln\left(\frac{\delta x\delta p}{\pi\hbar}\right),
\end{align}
where $1\geq\beta\geq 1/2$.

\section{Conclusions}

I have shown that quantum mechanical uncertainty relations for canonically conjugate variables can be expressed as inequalities involving the R\'enyi entropies. The simplicity of these relations indicates, in my opinion, that the R\'enyi entropy is an apt characteristic of the uncertainties in quantum measurements. A significant feature of the uncertainty relations (\ref{reur}), (\ref{seur}), (\ref{reurn}), and (\ref{angle}) is the appearance of the resolving power of the measuring apparatus. Since the R\'enyi entropy is an extension of the Shannon entropy, the new uncertainty relations generalize the entropic uncertainty relations derived before. The formulation of the uncertainty relations in terms of the R\'enyi entropies seems to indicate that a symmetrized version of the R\'enyi entropy (\ref{sym}) might be a useful concept.

\acknowledgments This research was supported by the Polish Ministry of Scientific Research Grant No. 008/P03/2003 Quantum Information and Quantum Engineering. I would like to thank Zofia Bialynicka-Birula and Karol \.Zyczkowski for discussions and Steeve Zozor for helpful correspondence.

\end{document}